\documentclass[conference,10pt]{IEEEtran}

\usepackage[dvips]{graphicx}
\usepackage{float}
\usepackage{times}
\usepackage{amsmath}
\usepackage{amsfonts}
\usepackage{gensymb}
\usepackage{cite}

\begin{document}
%
\title{A Fully Quaternion-Valued Capon Beamformer Based on Crossed-Dipole Arrays}

\author{\IEEEauthorblockN{Xiang Lan and Wei Liu\\}
\IEEEauthorblockA{Communications Research Group\\ Department of
Electronic and Electrical Engineering\\ University of Sheffield,
UK}}



%


\maketitle

\begin{abstract}
Quaternion models have been developed for both direction of arrival estimation and beamforming based on crossed-dipole arrays in the past. However, for almost all the models, especially for adaptive beamforming, the desired signal is still complex-valued and one example is the quaternion-Capon beamformer. However, since the complex-valued desired signal only has two components, while there are four components in a quaternion, only two components of the quaternion-valued beamformer output are used and the remaining two are simply removed. This leads to significant redundancy in its implementation. In this work, we consider a quaternion-valued desired signal and develop a full quaternion-valued  Capon beamformer, which has a better performance and a much lower complexity and is shown to be more robust against array pointing errors.


\end{abstract}


%
\IEEEpeerreviewmaketitle

\section{Introduction}
Electromagnetic (EM) vector sensor arrays can track the direction of arrival (DOA) of impinging signals as well as their polarization. A crossed-dipole sensor array, firstly introduced in \cite{compton81} for adaptive beamforming, works  by processing the received signals with a long polarization vector. Based on such a model, the beamforming problem is studied in detail in terms of output signal-to-interference-plus-noise ratio (SINR)~\cite{Nehorai_1999}. In \cite{godara1997,xu2004}, further detailed analysis was performed showing that the output SINR is affected by DOA and polarization differences.

Since there are four components for each vector sensor output in a crossed-dipole array, a quaternion model instead of long vectors has been adopted in the past for both adaptive beamforming and direction of arrival (DOA) estimation~\cite{miron2005,miron2006,gong2008,tao13a,liu14e}. In \cite{gou2011}, the well-known Capon beamformer was extended to the quaternion domain and a quaternion-valued Capon (Q-Capon) beamformer was proposed with the corresponding optimum solution derived.

However, in most of the beamforming studies, the signal of interest (SOI) is still complex-valued, i.e. with only two components: in-phase (I) and quadrature (Q). Since the output of a quaternion-valued beamformer is also quaternion-valued,  only two components of the quaternion are used to recover the SOI, which leads to redundancy in both calculation and data storage. However, with the development of quaternion-valued wireless communications~\cite{isaeva95a,wysocki09a,liu14n}, it is very likely that in the future we will have quaternion-valued signals as the SOI, where two traditional complex-valued signals with different polarisations arrive at the array with the same DOA. In such a case, a full quaternion-valued array model is needed to compactly represent the four-component desired signal and also make sure the four components of the quaternion-valued output of the beamformer  are fully utilised.  In this work, we develop such a model and propose a new quaternion-valued Capon beamformer, where both its input and output are quaternion-valued.


This paper is structured as follows. The full quaternion-valued array model is introduced in Section II and the proposed quaternion-valued Capon beamformer is developed in Section III. Simulation results are presented in IV, and conclusions are drawn in Section V.

\section{Quaternion model for array processing}\label{sec:QM}

A quaternion is constructed by four
components~\cite{hamilton66a,kantor89a}, with one real part and three imaginary parts, which is defined as $q = q_{a}+iq_{b}+jq_{c}+kq_{d}$, where $i, j, k$ are three different imaginary units and $ q_{a},q_{b},q_{c},q_{d}$ are real-valued. The multiplication
principle among such units is
 \begin{equation}
    i^{2} = j^{2} = k^{2} = ijk = -1,
\end{equation}
and
\begin{equation}
    ij = -ji = k, ki = -ik = j,jk = -ki =i
\end{equation}
The conjugate $q^*$ of $q$ is $q^* = q_{a}-iq_{b}-jq_{c}-kq_{d}$.

A quaternion number can be conveniently denoted as a combination of two complex numbers $q = c_{1}+ic_{2}$, where the complex number $c_{1}=q_{a}+jq_{c}$ and $c_{2}=q_{b}+jq_{d}$. We will use this form later to represent our quaternion-valued signal of interest.

\begin{figure}
  \centering
  \includegraphics[width=0.4\textwidth]{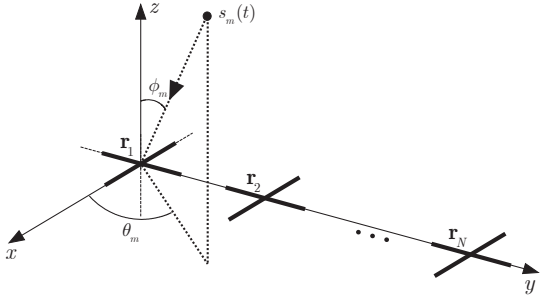}
  \caption{A crossed-dipole linear array with $N$ vector sensors.}\label{fig:01}
\end{figure}

Consider a uniform linear array with $N$ crossed-dipole sensors, as shown in Fig. \ref{fig:01}, where  the adjacent vector sensor spacing $d$ equals half wavelength, and the two components of each crossed-dipole are parallel to $x-$ and $y-$axes, respectively. A quaternion-valued narrowband signal $s_0(t)$ impinges upon the vector sensor array among other $M$ uncorrelated quaternion-valued interfering signals $\{s_{m}(t)\}_{m=1}^{M}$, with background noise $n(t)$. $s_0(t)$ can be decomposed into
\begin{equation}
s_0(t)=s_{01}(t)+is_{02}(t)\;,
\end{equation}
where $s_{01}(t)$ and $s_{02}(t)$ are two complex-valued sub-signals with the same DOA but different polarizations.

Assume that all signals are ellipse-polarized. The parameters, including DOA and polarization of the $m$-th signal are denoted by $(\theta_{m},\phi_m,\gamma_{m1},\eta_{m1})$ for the first sub-signal and $(\theta_{m},\phi_m,\gamma_{m2},\eta_{m2})$ for the second sub-signal. Each crossed-dipole sensor receives signals both in the $x$ and $y$ sub-arrays.

For signal $s_m(t)$, the corresponding received signals at the $x$ and $y$ sub-arrays are respectively given by~\cite{zhang2014}:
\begin{eqnarray}
    \emph{\textbf{x}}(t) = \emph{\textbf{a}}_{m1}p_{xm1}s_{m1}(t)+\emph{\textbf{a}}_{m2}p_{xm2}s_{m2}(t)\nonumber\\
    \emph{\textbf{y}}(t) = \emph{\textbf{a}}_{m1}p_{ym1}s_{m1}(t)+\emph{\textbf{a}}_{m2}p_{ym2}s_{m2}(t)
\end{eqnarray}
where $\emph{\textbf{x}}(t)$ represents the received part in the x-sub-array, $\emph{\textbf{y}}(t)$ represents the part in the y-sub-array, and  $(p_{xm1},p_{ym1})$ and $(p_{xm2},p_{ym2})$ are the polarizations of the two complex sub-signals in $x$ and $y$ directions, respectively, which are given by~\cite{hawes2015},
\begin{eqnarray}
    &p_{xm1}&=-\cos{\gamma_{m1}}\nonumber\\
    &p_{ym1}&=\cos{\phi_{m}}\sin{\gamma_{m1}}{e}^{j\eta_{m1}}\nonumber\\
    &p_{xm2}&=-\cos{\gamma_{m2}}\nonumber\\
    &p_{ym2}&=\cos{\phi_{m}}\sin{\gamma_{m2}}{e}^{j\eta_{m2}}, \quad\text{when}
    \quad \theta_{m}=\frac{\pi}{2}
\end{eqnarray}
Note that $\emph{\textbf{a}}_{m1}$ and $\emph{\textbf{a}}_{m2}$ are the steering vectors for the two sub-signals, which are equal to each other since the two sub-signals share the same DOA $(\theta_{m},\phi_m)$.
\begin{eqnarray}
    \emph{\textbf{a}}_{m1}=[1\quad e^{\frac{-j2\pi\sin{\theta_{m}}\sin{\phi_{m}}}
    {\lambda}},...e^{\frac{-j(N-1)2\pi\sin{\theta_{m}}\sin{\phi_{m}}}{\lambda}}]^{\text{T}}\nonumber\\
    \emph{\textbf{a}}_{m2}=[1\quad e^{\frac{-j2\pi\sin{\theta_{m}}\sin{\phi_{m}}}
    {\lambda}},...e^{\frac{-j(N-1)2\pi\sin{\theta_{m}}\sin{\phi_{m}}}{\lambda}}]^{\text{T}}
\end{eqnarray}
A quaternion model can be constructed by combining the two parts as below:
\begin{eqnarray}
    &\emph{\textbf{q}}_{m}(t)& = \emph{\textbf{x}}(t)+i\emph{\textbf{y}}(t)\\
    &&= \emph{\textbf{a}}_{m1}(p_{xm1}+ip_{ym1})s_{m1}(t)\nonumber\\
    &&+\emph{\textbf{a}}_{m2}(p_{xm2}+ip_{ym2})s_{m2}(t)\nonumber\\
    &&= \emph{\textbf{b}}_{m1}s_{m1}(t)+\emph{\textbf{b}}_{m2}s_{m2}(t)
\end{eqnarray}
where $\{\emph{\textbf{b}}_{m1}, \emph{\textbf{b}}_{m2}\}\in \mathbb H^{N \times 1}$ can be considered
as the composite quaternion-valued steering vector.
Combining all source signals and the noise together, the result is given by:
\begin{equation}
    \emph{\textbf{q}}(t) = \sum\limits^{M}_{m=0}(\emph{\textbf{b}}_{m1}s_{m1}(t)+\emph{\textbf{b}}_{m2}s_{m2}(t))+\emph{\textbf{n}}_{q}(t)
\end{equation}
where $\emph{\textbf{n}}_{q}(t) = \emph{\textbf{n}}_{x}(t)+i\emph{\textbf{n}}_{y}(t)$ is the quaternion-valued noise vector consisting of the two sub-array noise vectors $\emph{\textbf{n}}_{x}(t)$ and $\emph{\textbf{n}}_{y}(t)$.

\section{Full quaternion Capon beamformer}\label{sec:FQ}
\subsection{The Full Q-Capon Beamformer}\label{sub:MA}

To recover the SOI among interfering signals and noise, the basic idea is to keep
a unity response to the SOI at the beamformer output and then reduce the power/variance of the output as much as possible~\cite{capon69a,frost72a}.
The key to construct such a Capon beamformer in the quaternion domain is to design an appropriate constraint to make sure the quaternion-valued SOI can pass through the beamformer with the desired unity response.

Again note that the quaternion-valued SOI can be expressed as a combination of two complex sub-signals.  To construct such a constraint,  one choice is to make sure the first complex sub-signal of the SOI pass through the beamformer and appear in the real and $j$ components of the beamformer output, while the second complex sub-signal appear in the $i$ and $k$ components of the beamformer output.  Then, with a quaternion-valued weight vector \emph{\textbf{w}}, the constraint can be formulated as
\begin{equation}\label{eq:constraint}
    \emph{\textbf{w}}^{\text H}\emph{\textbf{C}} = \emph{\textbf{f}}
\end{equation}
where $\{\}^{\text H}$ is the Hermitian transpose (combination of the quaternion-valued conjugate and transpose operation), $\emph{\textbf{C}} = [\emph{\textbf{b}}_{01}\quad \emph{\textbf{b}}_{02}]$, and $\emph{\textbf{f}} =[1\quad i]$.

With this constraint, the beamformer output $z(t)$ is given by
\begin{eqnarray}\label{eq:output}\nonumber
    &z(t)& =\emph{\textbf{w}}^{\text H}\emph{\textbf{q}}(t)\nonumber\\
    &&= \underbrace{s_{01}(t)+is_{02}(t)}_{s_0(t)}+\emph{\textbf{w}}^{\text H}\emph{\textbf{n}}_{q}(t)\nonumber\\
    &&\quad+\sum\limits^{M}_{m=1}\emph{\textbf{w}}^{\text H}
    [\emph{\textbf{b}}_{m1}s_{m1}(t)+\emph{\textbf{b}}_{m2}s_{m2}(t)]
\end{eqnarray}
Clearly, the quaternion-valued SOI has been preserved at the output with the desired unity response.

Now, the full-quaternion Capon (full Q-Capon) beamformer can be formulated as
\begin{eqnarray}\label{eq:minrob}
    &\min&\emph{\textbf{w}}^{\text H}\emph{\textbf{R}}\emph{\textbf{w}} \text{   subject to  } \emph{\textbf{w}}^{\text H}\emph{\textbf{C}} = \emph{\textbf{f}}
\end{eqnarray}
where
\begin{equation}
\label{eq:QSM}
    \emph{\textbf{R}} = E\{\emph{\textbf{q}}(t)\emph{\textbf{q}}^{\text H}(t)\}\;.
\end{equation}

Applying the Lagrange multiplier method, we have
\begin{equation}
\label{eq:gradientw}
    l(\emph{\textbf{w}},\boldsymbol{\lambda})=\emph{\textbf{w}}^{\text H}\emph{\textbf{R}}\emph{\textbf{w}}+(\emph{\textbf{w}}^
    {\text H}\emph{\textbf{C}}-\emph{\textbf{f}})\boldsymbol{\lambda}^{\text H}
    +\boldsymbol{\lambda}(\emph{\textbf{C}}^{\text H}\emph{\textbf{w}}-\emph{\textbf{f}}^{\text H})
\end{equation}
where $\boldsymbol{\lambda}$ is a quaternion-valued vector.

The minimum can be obtained by setting the gradient of (\ref{eq:gradientw}) with respect
to $\emph{\textbf{w}}^{*}$ equal to a zero vector \cite{liu16c}. It is given by
\begin{equation}
    \nabla_{\emph{\textbf{w}}^{*}}l(\emph{\textbf{w}},\boldsymbol{\lambda}) =
    \frac{1}{2}\emph{\textbf{R}}\emph{\textbf{w}}+\frac{1}{2}\emph{\textbf{C}}\boldsymbol{\lambda}
    ^{\text H}= \emph{\textbf{0}}
\end{equation}

Considering all the constraints above, we obtain the optimum weight vector $\emph{\textbf{w}}_{opt}$ as follows
\begin{equation}
\label{eq:w}
    \emph{\textbf{w}}_{opt} = \emph{\textbf{R}}^{-1}\emph{\textbf{C}}(\emph{\textbf{C}}^{\text H}
    \emph{\textbf{R}}^{-1}\emph{\textbf{C}})^{-1}\emph{\textbf{f}}^{\text H}\;.
\end{equation}

A detailed derivation for the quaternion-valued optimum weight vector can be found at the Appendix.

In the next subsection, we give a brief analysis to show that by this optimum weight vector, the interference part at the beamformer output $z(t)$ in (\ref{eq:output}) has been suppressed effectively.
\subsection{Interference Suppression}\label{sub:Sup}

Expanding the covariance matrix, we have
\begin{equation}
    \emph{\textbf{R}} = E\{\emph{\textbf{q}}(t)\emph{\textbf{q}}^{\text H}(t)\}
    = \emph{\textbf{R}}_{i+n}+\sigma^2_1\emph{\textbf{b}}_{01}\emph{\textbf{b}}_{01}^{H}
    +\sigma^2_2\emph{\textbf{b}}_{02}\emph{\textbf{b}}_{02}^{\text H}
\end{equation}
where $\sigma^2_1,\sigma^2_2$ are the power of the two sub-signals of SOI and $\emph{\textbf{R}}_{i+n}$ denotes
the covariance matrix of interferences plus noise. 
Using the Sherman-Morrison formula, we then have
\begin{equation}
\label{eq:weight}
    \emph{\textbf{w}}_{opt} = \emph{\textbf{R}}_{i+n}^{-1}\emph{\textbf{C}}\boldsymbol{\beta}
\end{equation}
where $\boldsymbol{\beta} =(\emph{\textbf{C}}^{\text H}\emph{\textbf{R}}_{i+n}\emph{\textbf{C}})^{-1}
\emph{\textbf{f}}^{\text H}\in \mathbb H^{2 \times 1}$ is a quaternion vector.

Applying left eigendecomposition for quaternion matrix \cite{zhang1997,huang2001,le2004},
\begin{equation}
    \emph{\textbf{R}}_{i+n} = \sum\limits^{N}_{n=1}\alpha_{n}\emph{\textbf{u}}_{n}\emph{\textbf{u}}_{n}^{\text H}
\end{equation}
with  $\alpha_{1}\geq\text{...}\geq\alpha_{M-2}\textgreater {\alpha_{M-1}}=
\text{...}=\alpha_{N}=2\sigma^2_0  \in \mathbb R$, where $2\sigma^2_0$ denotes the noise power.

With sufficiently high interference to noise ratio (INR), the inverse of $\emph{\textbf{R}}_{i+n}$ can be approximated by
\begin{equation}
\label{eq:INV}
    \emph{\textbf{R}}_{i+n}^{-1} \approx \sum\limits^{N}_{n=M+1}\frac{1}{2\sigma^2_0}
    \emph{\textbf{u}}_{n}\emph{\textbf{u}}_{n}^{\text H}
\end{equation}
Then, we have
\begin{eqnarray}
    &\emph{\textbf{w}}_{opt}& = \sum\limits^{N}_{n=M+1}\frac{1}{2\sigma^2_0}
    \emph{\textbf{u}}_{n}\emph{\textbf{u}}_{n}^{\text H}\emph{\textbf{C}}\boldsymbol\beta =\sum\limits^{N}_{n=M+1}\emph{\textbf{u}}_{n}\rho_{n}
\end{eqnarray}
where $\rho_{n}$ is a quaternion-valued constant. Clearly, $\emph{\textbf{w}}_{opt}$ is the right linear combination of $\{u_{M+1},u_{M+2},\text{...},u_{N}\}$, and
$\emph{\textbf{w}}\in\text{$span_{R}$}\{u_{M+1},u_{M+2},\text{...},u_{N}\}$.

For those $M$ interfering signals, their quaternion steering vectors belong to
the space right-spanned by the related $M$ eigenvectors, i.e. $\emph{\textbf{b}}_{m1},\emph{\textbf{b}}_{m2}
\in \text{$span_{R}$}\{u_{1},u_{2},\text{...},u_{M}\}$. As a result,
\begin{eqnarray}
    \emph{\textbf{w}}_{opt}^{\text H}\emph{\textbf{b}}_{m1}\approx0,
    \emph{\textbf{w}}_{opt}^{\text H}\emph{\textbf{b}}_{m2}\approx0,
    m=1,4,\text{...},M
\end{eqnarray}
which shows that the beamformer has eliminated the interferences effectively.

\subsection{Complexity Analysis}\label{sub:COM}

In this section, we make a comparison of the computation complexity between the Q-Capon beamformer in \cite{gou2011} and our proposed full Q-Capon beamformer. To deal with a quaternion-valued signal, the Q-Capon beamformer has to process the two complex sub-signals separately to recover the desired signal completely, which means we need to apply the beamformer twice for a quaternion-valued SOI. However, for the full Q-Capon beamformer, the SOI is recovered directly by applying the beamformer once.

For the Q-Capon beamformer, the weight vector is calculated by $\textbf w=\emph{\textbf{R}}^{-1}\emph{\textbf{a}}_{0}(\emph{\textbf{a}}_{0}^{\text H}\emph{\textbf{R}}^{-1}\emph{\textbf{a}}_{0})^{-1}$, where $\textbf{a}_{0}$ is the steering vector for the complex-valued SOI. As an example, we use Gaussian elimination to calculate the matrix inversion $\emph{\textbf{R}}^{-1}$ and $\frac{1}{3}(N^{3}-N)$ quaternion-valued multiplications are needed, equivalent to $\frac{16}{3}(N^{3}-N)$ real-valued multiplications. Additionally, $\emph{\textbf{R}}^{-1} \emph{\textbf{a}}_{0}$ requires $16N^2$ real-valued multiplications, while $16(N^2+N)$ real multiplications are needed for $(\emph{\textbf{a}}_{0}^{\text H}\emph{\textbf{R}}^{-1}\emph{\textbf{a}}_{0})^{-1}$. In total, $\frac{16}{3}N^3+32N^2+\frac{80}{3}N$ real multiplications are needed. When processing a quaternion-valued signal, this number will be doubled and the total number of real multiplications becomes $\frac{32}{3}N^3+64N^2+\frac{160}{3}N$.

For the proposed full Q-Capon beamformer, in addition to calculating $\emph{\textbf{R}}^{-1}$, $32N^2$ real multiplications are required to calculate $\emph{\textbf{R}}^{-1}
\emph{\textbf{C}}$ and $32M^2+32M+96$ real multiplications for $(\emph{\textbf{C}}^{\text H}\emph{\textbf{R}}^{-1}\emph{\textbf{C}}_{0})^{-1}\emph{\textbf{f}}$.
In total, the number of real-valued multiplications is $\frac{16}{3}M^3+64M^2+\frac{272}{3}M+96$, which is roughly half of that of the Q-Capon beamformer.

\section{Simulations Results}\label{sec:sim}


In our simulations, we consider $10$ pairs of cross-dipoles with half wave-length spacing. All signals are assumed to arrive from the same plane of $\theta=90\degree$ and all interferences have the same polarization parameter $\gamma=60\degree$. For the SOI, the two sub-signals are
set to (90\degree, 1.5\degree, 90\degree, 45\degree) and (90\degree, 1.5\degree, 0\degree, 0\degree), with inferences coming from
(90\degree, 30\degree, 60\degree, -80\degree), (90\degree, -70\degree, 60\degree, 30\degree), (90\degree, -20\degree, 60\degree, 70\degree), (90\degree, 50\degree, 60\degree, -50\degree), respectively. The background noise is zero-mean quaternion-valued Gaussian. The power of SOI and all interfering signals
are set equal and SNR (INR) is 20dB.
\begin{figure}
  \centering
  \includegraphics[width=0.45\textwidth]{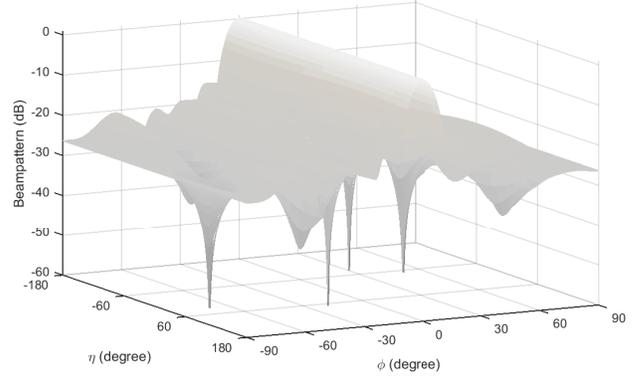}\\
  \caption{The resultant beam pattern with $\theta=90\degree$ and $\gamma=60\degree$. \label{fig:beampattern}}
  \vspace{3ex}
\end{figure}

Fig. \ref{fig:beampattern} shows the resultant 3-D beam pattern by the proposed beamformer, where the interfering signals from ($\phi,\eta$)=(30\degree, -80\degree),
(-70\degree, 30\degree), (-20\degree, 70\degree) and (50\degree, -50\degree) have all been effectively suppressed.

In the following, the output SINR performance of the two Capon beamformers (full Q-Capon and Q-Capon) is studied with the DOA and polarization
(90\degree, 1.5\degree, 90\degree, 45\degree) and (90\degree, 1.5\degree, 0\degree, 0\degree) for SOI and
(90\degree, 30\degree, 60\degree, -80\degree), (90\degree,- 70\degree, 60\degree, 30\degree), (90\degree, -20\degree, 60\degree, 70\degree),
 (90\degree, 50\degree, 60\degree, -50\degree) for interferences. Again, we have set SNR=INR=20dB. All results are obtained by averaging 1000 Monte-Carlo trials.

\begin{figure}[t]
  \centering
  \includegraphics[width=0.4\textwidth]{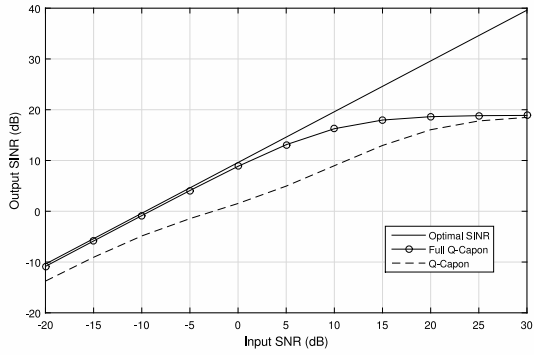}
  \caption{Output SINR versus input SNR (snapshots number 100).}\label{fig:SINR}
\end{figure}

Fig. \ref{fig:SINR} shows the output SINR performance versus SNR with $100$ snapshots, where the solid-line is for the optimal beamformer with infinite number of snapshots. For most of the input SNR range, in particular the lower range, the proposed full Q-Capon beamformer has a better performance than the Q-Capon beamformer. For very high input SNR values, these two beamformers have a very similar performance. 

\begin{figure}
  \centering
  \vspace{3ex}
  \includegraphics[width=0.4\textwidth]{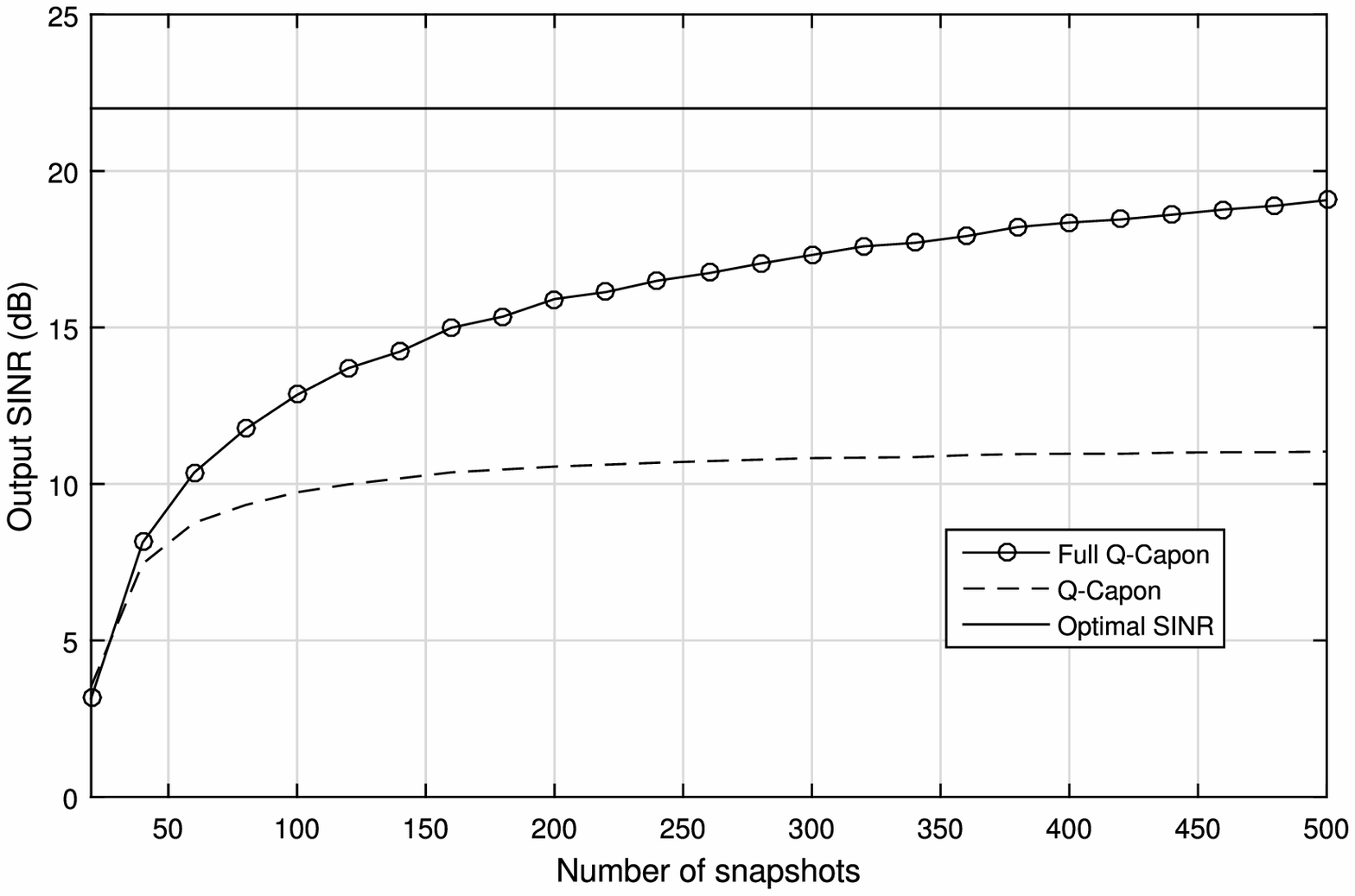}
  \caption{Output SINR versus snapshot number with SNR=SIR=15dB and 1\degree error. \label{fig:14}}
  \vspace{3ex}
\end{figure}

\begin{figure}
  \centering
  \includegraphics[width=0.4\textwidth]{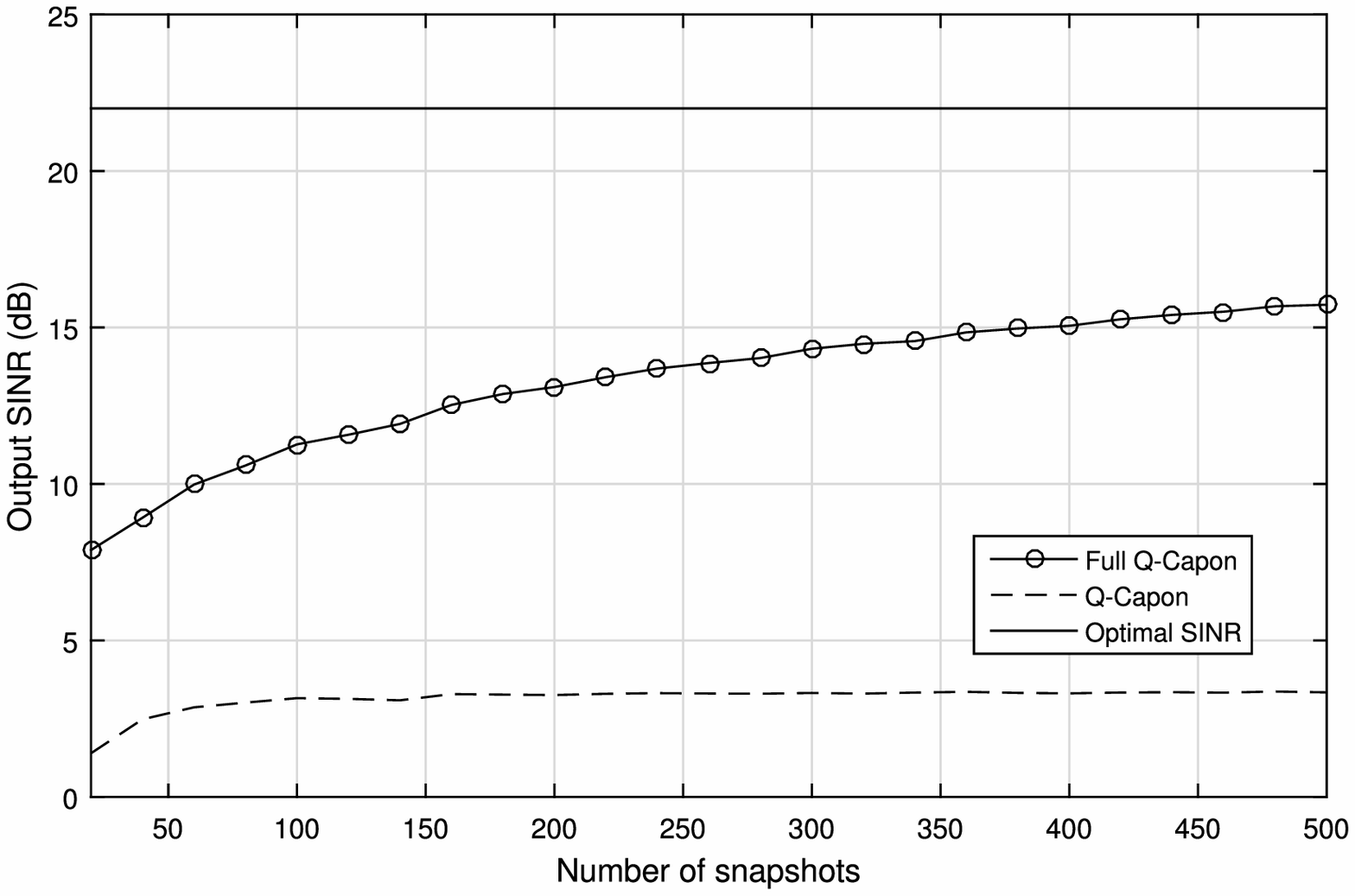}
  \caption{Output SINR versus snapshot number with SNR=SIR=15dB and 5\degree error. \label{fig:15}}
\end{figure}

Next, we investigate their performance in the presence of DOA and polarization errors. The  output SINR with respect to the number of snapshots is shown in Fig. \ref{fig:14} in the presence of $1\degree$ error for the SOI, where the real DOA and polarization parameters are (91\degree,2.5\degree,91\degree,46\degree) and (91\degree,2.5\degree,1\degree,1\degree). It can be seen that the full Q-Capon beamformer has achieved a much higher output SINR than the Q-Capon beamformer, and this gap increases with the increase of snapshot number.  Fig. \ref{fig:15} shows a similar trend in the presence of a $5\degree$ error. Overall, we can see that the proposed full Q-Capon beamformer is more robust against array pointing errors.

\section{Conclusions}\label{sec:con}
In this paper, a full quaternion model has been developed for adaptive beamforming based on crossed-dipole arrays, with a new full quaternion Capon beamformer proposed. Different from previous studies in quaternion-valued adaptive beamforming, we have considered a quaternion-valued desired signal, given the recent development in quaternion-valued wireless communications research. The proposed beamformer has a better performance and a much lower computational complexity than a previously proposed Q-Capon beamformer and is also shown to be more robust against array pointing errors, as demonstrated by computer simulations. 

\begin{appendix}

The gradient of a quaternion vector $\textbf{u}=\emph{\textbf{w}}^{\text H}\emph{\textbf{C}}\boldsymbol{\lambda}^
{\text H}$ with respect to $\emph{\textbf{w}}^{*}$
can be calculated as below:
\begin{eqnarray}
\label{eq:gra}
    &\nabla_{\emph{\textbf{w}}^{*}}{\textbf{u}}&
    = [\nabla_{\emph{{w}}_{1}^{*}}\textbf{u} \quad \nabla{\emph{{w}}_{2}^{*}}\textbf{u}\quad \text{...} \nabla_{\emph{{w}}_{n}^{*}}\textbf{u}]^{T}
\end{eqnarray}
where $w_n$, $n=1, 2, \cdots, N$ is the $n$-th quaternion-valued coefficient of the beamformer. Then,
\begin{eqnarray}
    &\nabla_{\emph{{w}}_{1}^{*}}{\textbf{u}}& = \frac{1}{4}(\nabla_{\emph{{w}}_{1a}}{\textbf{u}}+\nabla_{\emph{{w}}_{1b}}{\textbf{u}}i+\nabla_{\emph{{w}}_{1c}}\textbf{u}j+\nabla_{\emph{{w}}_{1d}}\textbf{u}k)
\end{eqnarray}
where
\begin{equation}
    \emph{{w}}_{1}^{*} = \emph{{w}}_{1a}-\emph{{w}}_{1b}i-\emph{{w}}_{1c}j-\emph{{w}}_{1d}k
\end{equation}

Since $\emph{{w}}_{1a}$ is real-valued, with the chain rule~\cite{liu16c}, we have
\begin{eqnarray}
    &\nabla_{\emph{{w}}_{1a}}\textbf{u}& =\nabla_{\emph{{w}}_{1a}}({\emph{\textbf{w}}^{\text H}})
    \emph{\textbf{C}}\boldsymbol{\lambda}^{\text H}+\emph{\textbf{w}}^{\text H}\nabla_{\emph{{w}}_{1a}}({\emph{\textbf{C}}\boldsymbol{\lambda}^{\text H}})\nonumber\\
    && =[1\quad 0\quad 0\quad\text{...}\quad 0]\emph{\textbf{C}}\boldsymbol{\lambda}^{\text H}
\end{eqnarray}
Similarly,
\begin{eqnarray}
    &\nabla_{\emph{{w}}_{1b}}\textbf{u}& =[-i\quad 0\quad 0\quad\text{...}\quad 0]
    \emph{\textbf{C}}\boldsymbol{\lambda}^{\text H}\nonumber\\
    &\nabla_{\emph{{w}}_{1c}}\textbf{u}& =[-j\quad 0\quad 0\quad\text{...}\quad 0]
    \emph{\textbf{C}}\boldsymbol{\lambda}^{\text H}\nonumber\\
    &\nabla_{\emph{{w}}_{1d}}\textbf{u}& =[-k\quad 0\quad 0\quad\text{...}\quad 0]
    \emph{\textbf{C}}\boldsymbol{\lambda}^{\text H}\nonumber
\end{eqnarray}
Hence,
\begin{eqnarray}
    &\nabla_{\emph{{w}}_{1}^{*}}\textbf{u}& = \frac{1}{4}(4{\text {Real}}(\emph{\textbf{C}}\boldsymbol{\lambda}^{\text H})_{1})={\text {Real}}(\emph{\textbf{C}}\boldsymbol{\lambda}^{\text H})_{1}\;,
\end{eqnarray}
where the subscript $\{\}_{1}$ in the last item means taking the first entry of the vector.

Finally,
\begin{equation}
\label{eq:w1}
    \nabla_{\emph{\textbf{w}}^{*}}\textbf{u}={\text {Real}}(\emph{\textbf{C}}\boldsymbol{\lambda}^{\text H})
\end{equation}

The gradient of the quaternion vector $\textbf{v}=\boldsymbol{\lambda}\emph{\textbf{C}}^{\text H}\emph{\textbf{w}}$ with respect to $\emph{\textbf{w}}^{*}$ can be calculated in the same way:
\begin{eqnarray}
    &\nabla_{\emph{{w}}_{1a}}\textbf{v}& =\boldsymbol{\lambda}\emph{\textbf{C}}^{\text H}\nabla_{\emph{{w}}_{1a}}\emph{\textbf{w}}
    +\nabla_{\emph{{w}}_{1a}}(\boldsymbol{\lambda}\emph{\textbf{C}}^{\text H})
    \emph{\textbf{w}}
    \nonumber\\
    && =\boldsymbol{\lambda}\emph{\textbf{C}}^{\text H}[1 \quad 0\quad 0 \quad\text{...}\quad 0]^{\text T}
\end{eqnarray}
Similarly,
\begin{eqnarray}
    &\nabla_{\emph{{w}}_{1b}}\textbf{v}& =\boldsymbol{\lambda}\emph{\textbf{C}}^{\text H}[i \quad 0
    \quad 0 \quad\text{...}\quad 0]^{\text T}\nonumber\\
    &\nabla_{\emph{{w}}_{1c}}\textbf{v}& =\boldsymbol{\lambda}\emph{\textbf{C}}^{\text H}[j \quad 0
    \quad 0 \quad\text{...}\quad 0]^{\text T}\nonumber\\
    &\nabla_{\emph{{w}}_{1d}}\textbf{v}& =\boldsymbol{\lambda}\emph{\textbf{C}}^{\text H}[k \quad 0
    \quad 0 \quad\text{...}\quad 0]^{\text T}
\end{eqnarray}
Thus, the gradient can be expressed as
\begin{eqnarray}
    &\nabla_{\emph{{w}}_{1}^{*}}\textbf{v}&=-\frac{1}{2}(\emph{\textbf{C}}\boldsymbol{\lambda}^{\text H})_{1}^{*}
\end{eqnarray}
Finally,
\begin{equation}
\label{eq:w2}
    \nabla_{\emph{\textbf{w}}^{*}}\textbf{v}=-\frac{1}{2}(\emph{\textbf{C}}\boldsymbol{\lambda}^{\text H})^{*}
\end{equation}

The gradient of $c_{\textbf{w}} = {\emph{\textbf{w}}^{\text H}\emph{\textbf{R}}\emph{\textbf{w}}}$ can be
calculated as follows.
\begin{equation}
    \nabla_{\emph{\textbf{w}}^{*}}c_{\textbf{w}} =
    [\nabla_{\emph{{w}}_{1}^{*}}c_{\textbf{w}}\quad \nabla_{\emph{{w}}_{2}^{*}}c_{\textbf{w}}\quad
    \text{...}\quad \nabla_{\emph{{w}}_{n}^{*}}c_{\textbf{w}}]^{\text T}
\end{equation}
\begin{equation}
    \nabla_{\emph{{w}}_{1}^{*}}c_{\textbf{w}} =\frac{1}{4}(\nabla_{\emph{{w}}_{1a}}c_{\textbf{w}}+\nabla_{\emph{{w}}_{1b}}c_{\textbf{w}}i
    +\nabla_{\emph{{w}}_{1c}}c_{\textbf{w}}j+\nabla_{\emph{{w}}_{1d}}c_{\textbf{w}}k)
\end{equation}

Now we calculate the gradient of $c_{\textbf{w}}$ with respect to the four components of $\emph{{w}}_{1}$.
\begin{eqnarray}
    &\nabla_{\emph{{w}}_{1a}}c_{\textbf{w}}& =\nabla_{\emph{{w}}_{1a}}({\emph{\textbf{w}}^{\text H}\emph{\textbf{R}}})\emph{\textbf{w}}+\emph{\textbf{w}}^{\text H}\emph{\textbf{R}}\nabla_{\emph{{w}}_{1a}}{\emph{\textbf{w}}}\nonumber\\
    && = [1 \quad 0 \quad 0 \quad \text{...} \quad 0]\emph{\textbf{R}}\emph{\textbf{w}}\nonumber\\
    &&\quad +\emph{\textbf{w}}^{\text H}\emph{\textbf{R}}[1 \quad 0 \quad 0 \quad \text{...} \quad 0]^{\text T}
\end{eqnarray}
The other three components are,
\begin{eqnarray}
    &\nabla_{\emph{{w}}_{1b}}c_{\textbf{w}}& = [-i \quad 0 \quad 0 \quad \text{...} \quad 0]\emph{\textbf{R}}\emph{\textbf{w}}\nonumber\\
    &&\quad +\emph{\textbf{w}}^{\text H}\emph{\textbf{R}}[i \quad 0 \quad 0 \quad \text{...}
    \quad 0]^{\text T}\nonumber\\
    &\nabla_{\emph{{w}}_{1c}}c_{\textbf{w}}& = [-j \quad 0 \quad 0 \quad \text{...} \quad 0]\emph{\textbf{R}}\emph{\textbf{w}}\nonumber\\
    &&\quad +\emph{\textbf{w}}^{\text H}\emph{\textbf{R}}[j \quad 0 \quad 0 \quad \text{...}
    \quad 0]^{\text T}\nonumber\\
    &\nabla_{\emph{{w}}_{1d}}c_{\textbf{w}}& = [-k \quad 0 \quad 0 \quad \text{...} \quad 0]\emph{\textbf{R}}\emph{\textbf{w}}\nonumber\\
    &&\quad +\emph{{w}}^{\text H}\emph{\textbf{R}}[k \quad 0 \quad 0 \quad \text{...}
    \quad 0]^{\text T}\nonumber
\end{eqnarray}
Hence,
\begin{eqnarray}
    &\nabla_{\emph{{w}}_{1}^{*}}c_{\textbf{w}}&={\text {Real}}(\emph{\textbf{R}}\emph{\textbf{w}})_{1}-
    \frac{1}{2}(\emph{\textbf{R}}\emph{\textbf{w}})_{1}^{*}=\frac{1}{2}(\emph{\textbf{R}}\emph{\textbf{w}})_{1}
\end{eqnarray}
Finally,
\begin{equation}
\label{eq:w3}
    \nabla_{\emph{\textbf{w}}^{*}}c_{\textbf{w}}=\frac{1}{2}\emph{\textbf{R}}\emph{\textbf{w}}
\end{equation}

Combining (\ref{eq:w1}), (\ref{eq:w2}) and (\ref{eq:w3}), with (\ref{eq:gradientw}), we have
\begin{equation}
    \nabla_{\emph{\textbf{w}}^{*}}l(\emph{\textbf{w}},\boldsymbol{\lambda}) =\frac{1}{2}
    (\emph{\textbf{R}}\emph{\textbf{w}}+\emph{\textbf{C}}\boldsymbol{\lambda}^{\text H})
    = {\textbf{0}}
\end{equation}
Further,
\begin{equation}
\label{eq:35}
    \emph{\textbf{w}}=-2\emph{\textbf{R}}^{-1}\emph{\textbf{C}}\boldsymbol{\lambda}^{\text H}
\end{equation}
Subsituting (\ref{eq:35}) into (\ref{eq:constraint}),
\begin{eqnarray}
    \boldsymbol{\lambda} = -\frac{1}{2}\emph{\textbf{f}}(\emph{\textbf{C}}^{\text H}\emph{\textbf{R}}^{-1}
    \emph{\textbf{C}})^{-1}
\end{eqnarray}
Finally,
\begin{equation}
    \emph{\textbf{w}}=\emph{\textbf{R}}^{-1}\emph{\textbf{C}}(\emph{\textbf{C}}^{\text H}\emph{\textbf{R}}^{-1}
    \emph{\textbf{C}})^{-1}\emph{\textbf{f}}^{\text H}
\end{equation}

\end{appendix}

\bibliographystyle{IEEEtran}
\bibliography{mybib}

\end{document}